\documentclass{article}

\usepackage{arxiv}

\usepackage[utf8]{inputenc}
\usepackage[T1]{fontenc}
\usepackage[english]{babel}
\usepackage{amsmath}
\usepackage{amsfonts}
\usepackage{graphicx}
\usepackage{subcaption}
\usepackage{booktabs}
\usepackage{multirow}
\usepackage{url}
\usepackage{microtype}
\usepackage[colorlinks=true, allcolors=blue]{hyperref}

\usepackage{authblk}

\title{Diagnostically Meaningful, Privacy-aware Synthetic Chest CT via Class-Conditional Latent Sampling in Diffusion Autoencoders}
\renewcommand{\shorttitle}{Privacy-aware Synthetic Chest CT via Diffusion Autoencoders}

\author[1,2,3,4,*,+]{Johannes B. Thalhammer}
\author[1,2,*]{Kacper Ogórek}
\author[1,2,3]{Tina Dorosti}
\author[1,2]{Sebastian Peterhansl}
\author[1,2]{Daniel Frey}
\author[1,2]{Zijin Huang}
\author[1,2]{Theresa Hiu}
\author[2,5]{Tobias Lasser}
\author[1,2,3,4]{Daniela Pfeiffer}
\author[1,2,3,4]{Franz Pfeiffer}
\author[1,2]{Florian Schaff}
\affil[1]{Chair of Biomedical Physics, Department of Physics, School of Natural Sciences, Technical University of Munich, 85748 Garching, Germany}
\affil[2]{Munich Institute of Biomedical Engineering, Technical University of Munich, 85748 Garching, Germany}
\affil[3]{Department of Diagnostic and Interventional Radiology, School of Medicine and Health, Klinikum rechts der Isar, Technical University of Munich, 81675 Munich, Germany}
\affil[4]{Institute for Advanced Study, Technical University of Munich, 85748 Garching, Germany}
\affil[5]{Computational Imaging and Inverse Problems, Department of Computer Science, School of Computation, Information, and Technology, Technical University of Munich, 85748 Garching, Germany}
\affil[*]{Contributed equally}
\affil[+]{Corresponding author: \texttt{johannes.thalhammer@tum.de}}

\begin{document}
\maketitle
\begin{abstract}
Large, well-annotated datasets are essential for developing medical artificial intelligence models, yet data sharing is often restricted by privacy regulations.
Synthetic images may help address this limitation if they preserve diagnostic information while reducing the risk of reproducing patient data.
We developed a diffusion autoencoder-based framework for class-conditional synthetic chest CT generation that incorporates latent-space distance filtering to reduce the likelihood of regenerating training samples.

A diffusion autoencoder was trained on the public RSNA 2020 Pulmonary Embolism (PE) dataset. PE-positive and PE-negative images were synthesized by sampling latent representations from class-specific kernel density estimates. Samples with a nearest-neighbor latent-space distance below a predefined threshold were rejected. To assess diagnostic utility, ResNet-50 classifiers were trained across 10 independent runs using real, synthetic, or mixed real--synthetic datasets. Performance was evaluated on a held-out real test set using the area under the receiver operating characteristic curve (AUC), with DeLong 95\% confidence intervals (CIs).

Models trained on synthetic data achieved an AUC of 0.862 (95\% CI, 0.851--0.874), only slightly lower than models trained on real data (AUC = 0.872, 95\% CI, 0.861--0.883) and mixed real--synthetic data (AUC = 0.865, 95\% CI, 0.855--0.876). These differences were statistically significant ($p < 0.05$) but small in absolute terms. A nearest-neighbor copy-detection analysis of 38,000 synthetic slices found no direct copies of real training images.

These findings indicate that the synthetic CT images preserve diagnostically meaningful information for image-level PE classification while incorporating measures intended to enhance privacy awareness and reduce similarity to real patient data.

\end{abstract}

\section*{Introduction}

Artificial intelligence (AI) methods are increasingly investigated for applications throughout the medical imaging workflow, including clinical education, image interpretation, and treatment planning \cite{najjar2023redefining, duong2019artificial, ghaffarnia2023evaluation, 10.1007/978-3-030-87202-1_51}.
The development and evaluation of these methods depend on access to large and diverse labeled imaging datasets.
However, medical images are costly to acquire and annotate, expert labeling is time-consuming, and sharing of medical images across institutions is restricted by patient privacy requirements and data protection regulations.
Two complementary strategies have been proposed to address this restricted access. Federated learning enables decentralized model training without transferring the underlying datasets \cite{long2021federated}, yet requires institutions to host and maintain local infrastructure for every collaborative study, and does not inherently prevent a trained model from memorizing sensitive information \cite{liu2022threats}.
Synthetic medical images offer a different strategy: generative models can learn the distribution represented in an existing dataset and produce new samples that preserve information relevant to a downstream diagnostic task \cite{morrison2021generative, M_ller_Franzes_2023}, potentially providing a source of training data that does not correspond to any single real patient.
Unlike federated learning, synthetic datasets can be accessed and shared directly, making them particularly useful when images are needed for model development, methodological evaluation, or collaboration outside a controlled federated setting.
Before such applications can be considered, however, the extent to which synthetic images retain diagnostically relevant information must be established empirically. If classifiers trained exclusively on synthetic images perform comparably to those trained on real images when evaluated on held-out real data, synthetic datasets could support model development without requiring direct access to the corresponding real patient data.

Denoising diffusion probabilistic models have emerged as a widely used class of generative models \cite{ho2020denoising}. Unlike generative adversarial networks, which are often difficult to train, DDPMs avoid adversarial optimization and can generate high-fidelity images \cite{dhariwal2021diffusion, betker2023improving}, motivating their investigation for medical image generation and related applications \cite{M_ller_Franzes_2023, dorjsembe2023meddiffusion}. Diffusion autoencoders extend this framework by combining a semantic encoder with a conditional diffusion model: the semantic encoder maps an image to a compact latent representation containing high-level image information, while the diffusion model generates an image conditioned on this representation. This structure enables image generation to be controlled directly through the semantic latent space, offering a natural mechanism for class-conditional synthesis.

This latent-space structure suggests a practical workflow for institutions holding proprietary, labeled imaging data that cannot be freely shared. A diffusion autoencoder can first be pretrained once, in a fully label-free manner, on a large corpus of unlabeled images. An institution can then apply the resulting semantic encoder to its own proprietary, labeled dataset, extracting a compact latent vector for each image together with its associated label.
Class-conditional distributions can be estimated directly from these labeled latent vectors, for instance via kernel density estimation (KDE), and new latent vectors sampled from these distributions can be decoded back into images. The resulting synthetic, labeled dataset is derived from, but not identical to, the original proprietary data, and could in principle be shared or used for downstream model development in place of the original images.

\begin{figure}[ht]
	\centering
	\includegraphics[width=1.0\textwidth]{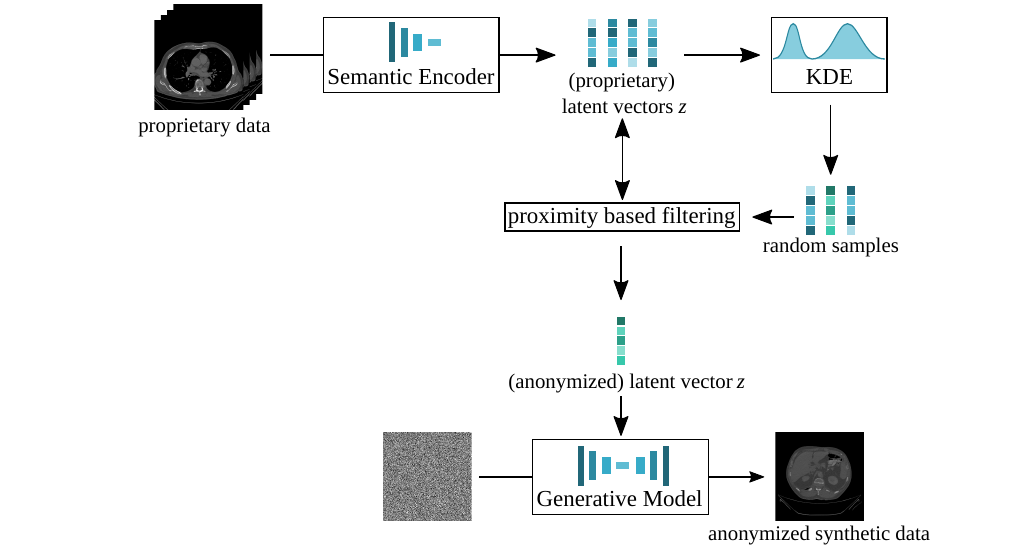}
    \caption{Overview of the proposed framework for generating privacy-aware, class-conditional synthetic images from proprietary data. A diffusion autoencoder, pretrained on a large unlabeled image corpus, provides a semantic encoder that extracts a latent vector for each image in a proprietary, labeled dataset. Distributions estimated from these latent vectors via kernel density estimation (KDE) enable sampling of new latent vectors, which are decoded into synthetic images. A filtering step rejects sampled vectors too close to real proprietary vectors, reducing the risk of regenerating real images. The pipeline is applied separately to each class (PE-positive and PE-negative).}
	\label{fig:pipeline}
\end{figure}

In this study, we develop and evaluate such a framework for generating class-specific chest CT images. The diffusion autoencoder is first trained without class labels on a large set of images, jointly learning a semantic encoder and a latent-conditioned diffusion model. Labeled CT images are then encoded into the semantic latent space, and separate distributions are estimated for pulmonary embolism (PE)-positive and PE-negative images using KDE. New latent vectors sampled from these class-conditional distributions are subsequently decoded to generate synthetic images.
To limit the generation of samples resembling real images, we additionally apply a latent-distance rejection procedure: each sampled latent vector is compared with the latent representations of the real training images, and samples falling below a predefined nearest-neighbor distance threshold are rejected. We refer to this practical safeguard as \emph{privacy-aware} generation, as distinct from methods that provide a formal privacy guarantee. An overview of the framework is shown in Fig.~\ref{fig:pipeline}.

We evaluate the framework in four steps. First, we examine whether the learned latent representations encode information associated with PE status, and whether this information is preserved after KDE-based sampling.
Second, we assess the similarity between real and synthetic images and evaluate the diagnostic utility of the generated data by comparing image-level PE classifiers trained exclusively on either real or synthetic images and tested on held-out real images. As a secondary experiment, we evaluate whether combining real and synthetic images in a class-balanced training set improves classifier performance beyond real-only training.
Third, we compare Grad-CAM saliency maps qualitatively across training conditions to examine which image regions the resulting classifiers rely on.
Finally, we perform a copy-detection analysis to assess whether synthetic images closely reproduce individual real training images, as a practical check on patient privacy.

\section*{Related work}
Several recent studies have applied diffusion models to class- or condition-specific medical image synthesis and related downstream tasks. Dorjsembe et al. \cite{dorjsembe2023meddiffusion} proposed Med-DDPM, a diffusion model for 3D brain MRI synthesis conditioned on segmentation masks via channel-wise concatenation with the network input.
Atad et al. \cite{atad2024counterfactual} used the semantic latent space of a diffusion autoencoder to generate counterfactual explanations for medical image classification and regression, without requiring a separate explanation model.
Munia and Imran \cite{munia2025classndiff} introduced Class-N-Diff, a latent diffusion model with an integrated classifier that guides class-conditional generation of dermoscopic skin images to improve both image fidelity and diagnostic fairness.
Cahan et al. \cite{cahan2024xray2ctpa} proposed X-ray2CTPA, which generates synthetic 3D CT pulmonary angiography volumes from 2D chest radiographs using a conditional diffusion model and uses the resulting volumes to improve pulmonary embolism classification.

To preserve patient privacy, synthetic data should capture the relevant characteristics of the original data without reproducing individual training samples. Privacy-preserving generative models have been studied broadly. Padariya et al. \cite{padariya2025privacy} provide an overview, categorizing approaches by their underlying privacy mechanisms and highlighting the lack of a single adequate privacy or utility metric across settings. Differential privacy \cite{dwork2014algorithmic} offers formal guarantees and can be enforced during the training of diffusion models \cite{abadi2016deep, dockhorn2023differentially}. However, it typically degrades image quality, also in medical image synthesis \cite{daum2024differentially}, and it protects only the data the generative model was trained on, not the images from which the conditioning latent vectors are derived, which in our framework are precisely the proprietary data. Empirical approaches instead search generated images for replicated training samples \cite{carlini2023extracting, somepalli2023diffusion} or mitigate such copying \cite{somepalli2023understanding}. For medical diffusion models specifically, Dar et al. \cite{dar2023memorization} showed that 3D latent diffusion models trained on coronary CT angiography and knee MRI data can reproduce their training scans, identifying such memorized samples with a self-supervised contrastive detector, and in a follow-up study \cite{dar2024unconditional} found that memorization in unconditional latent diffusion models trained on CT, MR, and X-ray data is common but can be reduced through data augmentation and larger training sets. We use the term \emph{privacy-aware} for this practical category of safeguard: our framework reduces the risk of regenerating real images through its latent-distance rejection mechanism, but this does not amount to a formal privacy guarantee.

\section*{Methods}

\subsection*{Datasets}
\begin{figure}
	\centering
	\includegraphics[width=1.0\textwidth]{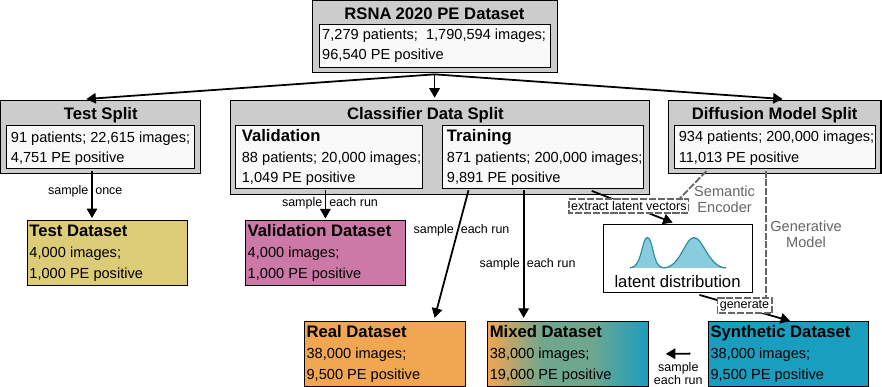}
	\caption{Data flowchart of the RSNA 2020 PE dataset. The data was split on patient level in test, validation, classifier training, and diffusion model training split. The mixed dataset was sampled from 50\% synthetic and 50\% real data. Each random sampling process was redone for each of the 10 classifier training runs.}
	\label{fig:class_data}
\end{figure}

This retrospective study used two clinical datasets, both publicly available and de-identified: the RSNA 2020 PE CT dataset \cite{RSNA-PE} and the Stanford Multimodal Pulmonary Embolism Dataset (RadFusion) \cite{stanford_aimi_radfusion}.
The RSNA dataset includes CT pulmonary angiography scans acquired on scanners from multiple manufacturers and collected from multiple institutions. The RadFusion dataset was collected at Stanford University Medical Center and comprises scans performed under the pulmonary embolism protocol.
Both datasets provide binary image-level labels indicating the presence of acute PE on each CT slice. For the RSNA dataset, labels were established by consensus among multiple board-certified radiologists with subspecialty training in thoracic imaging \cite{RSNA-PE}. In the RadFusion dataset, labels were assigned by two board-certified radiologists and verified by a senior radiologist \cite{stanford_aimi_radfusion}. Demographic information, such as age and sex, was not used, as the study focused exclusively on image-level classification.

Figure \ref{fig:class_data} summarizes the RSNA 2020 PE dataset and the patient-level disjoint splits used for diffusion model training, classifier training, and evaluation. We used a subset of the full RSNA 2020 PE dataset, sized to balance compute and training time for the large number of experiments in this study against providing sufficient data for stable density estimation and well-powered classifier comparisons.

The \textbf{Diffusion Model Split} contained 934 patients (200,000 slices; 11,013 PE-positive) and was used to train the diffusion autoencoder.

The \textbf{Classifier Data Split} consisted of a training subset (871 patients; 200,000 slices; 9,891 PE-positive) and a validation subset (88 patients; 20,000 slices; 1,049 PE-positive). From the training subset, three datasets were constructed for classifier training. \textbf{Real Datasets} (38,000 images; 9,500 PE-positive) were sampled directly from the training set for each training run. A \textbf{Synthetic Dataset} (38,000 images; 9,500 PE-positive) was generated using a class-conditional KDE fitted to latent vectors from the training set and sampled through the diffusion model. \textbf{Mixed Datasets} (38,000 images; 19,000 PE-positive) were created by sampling 9,500 PE-positive and 9,500 PE-negative images from each of the real and synthetic pools, yielding 19,000 PE-positive and 19,000 PE-negative images (38,000 total) with equal real/synthetic contribution within each class. \textbf{Validation Datasets} with 4,000 real images (1,000 PE-positive) were sampled for each run (real, synthetic and mixed) from the validation split.

Final evaluation was performed on a fixed \textbf{Test Dataset} (4,000 images; 1,000 PE-positive) sampled from a separate split of 91 patients (22,615 slices; 4,751 PE-positive).

The RadFusion dataset \cite{stanford_aimi_radfusion} was partitioned analogously to the RSNA 2020 PE dataset, except that no dedicated diffusion model split was created, as the diffusion model was trained exclusively on the RSNA 2020 PE dataset. Accordingly, the RadFusion dataset serves as an external dataset for evaluating the robustness of the generative framework. The training split comprised 1,454 patients and 572,318 images, including 39,800 PE-positive cases. The validation split contained 193 patients and 75,118 images (2,994 PE-positive), and the test split contained 190 patients and 75,142 images (3,123 PE-positive). The \textbf{Test}, \textbf{Validation}, \textbf{Real}, \textbf{Mixed}, and \textbf{Synthetic Datasets} were generated using the same sampling procedure as for RSNA 2020 PE, with identical numbers of images and PE-positive cases in each instance.

\subsection*{Diffusion Autoencoder framework \& training}

We use a Diffusion Autoencoder \cite{preechakul2022diffusion} framework based on denoising diffusion probabilistic models (DDPMs) \cite{ho2020denoising} to generate synthetic chest CT images. The framework consists of two main components, both operating on single-channel 256$\times$256 pixel inputs.

The first component is a semantic encoder $E$, which maps an input image $x_0$ to a 512-dimensional latent representation vector, $z = E(x_0)$. This latent vector captures high-level semantic and anatomical information and serves as a compact representation of the input image. During image synthesis, the latent vector guides the generation process towards anatomically realistic chest CT slices.

The second component is a diffusion-based generative model using a U-Net architecture as a backbone. During training, the U-Net $\epsilon_{\theta}$ learns to predict the noise $\epsilon$ added during the forward diffusion process. During inference, images are synthesized by iteratively reversing the diffusion process from $t=T$ to $t=0$, starting from a Gaussian noise sample $x_T \sim \mathcal{N}(0,I)$. The reverse process is conditioned on the diffusion timestep $t$, which specifies the current noise level, and optionally on the semantic latent vector $z$.

Both the semantic encoder and the U-Net are constructed from the same residual block comprising three consecutive residual sub-blocks of GroupNorm, Convolution, SiLU, Convolution, followed by a sigmoidal self-gating unit, inspired by \cite{dauphin2017languagemodelinggatedconvolutional, oktay2018attentionunetlearninglook}. The semantic encoder is a purely contractive network with five resolution levels and feature dimensions [32, 64, 128, 256, 512], where each level contains a single residual block followed by MaxPool downsampling. The resulting feature map is flattened and linearly projected to yield the 512-dimensional semantic latent vector $z$. The diffusion U-Net adopts a symmetric encoder-decoder architecture with four resolution levels of feature dimensions [64, 128, 256, 512] and a 1024-channel bottleneck. Each encoder and decoder level likewise consists of a single residual block, with MaxPool downsampling and bilinear upsampling, respectively. Following each residual block, an embedding injection module additively injects conditioning signals into the feature maps: the diffusion timestep
$t$ is encoded via sinusoidal positional encoding and summed with the semantic latent $z$. This joint embedding is processed by a dense layer, projected to the dimensionality of the corresponding feature maps, and added to them. Skip connections link encoder and decoder pathways by concatenating feature maps at corresponding resolution levels. A schematic illustration of the architecture is provided in the Supplementary Material.

Based on \cite{preechakul2022diffusion, song2022denoisingdiffusionimplicitmodels}, the distortion process is based on signal-ratio $\sqrt{\bar{\alpha}_t}$, following cosine scaling, and noise-ratio $\sqrt{1-\bar{\alpha}_t}$ schedulers defining image degradation during the forward diffusion process:

\begin{equation}
    x_t = \sqrt{\bar{\alpha}_t}\, x_0 +  \sqrt{1-\bar{\alpha}_t}\, \epsilon, \qquad \epsilon \sim \mathcal{N}(0, I).
    \label{eq:distortion}
\end{equation}

To improve training stability and facilitate high-resolution image synthesis, a deterministic frequency-domain low-pass filter $f(t)$ inspired by \cite{hoogeboom2024blurringdiffusionmodels} is additionally applied to $x_t$, progressively removing high-frequency image content during image degradation. The exact formulas for $\sqrt{\bar{\alpha}_t}$ and $f(t)$ are given in the Supplementary Material. The model is trained by minimizing

\begin{equation}
    L = \mathbb{E}_{x_0,\, \epsilon, t}
    \left\| \epsilon - \epsilon_{\theta}\big(x_t,\, t,\, z\big) \right\|_2^2,
    \label{eq:loss}
\end{equation}

where $\epsilon$ denotes the sampled Gaussian noise and $t \sim U(0,1)$ is a uniformly sampled diffusion timestep. As in standard DDPM training, only a single randomly sampled diffusion step is optimized per training iteration rather than explicitly reversing the entire diffusion chain. To enable classifier-free guidance during sampling, $z$ was replaced by the null (unconditional) vector $\emptyset$ for 10\% of the training samples, so that the same model also learns unconditional generation.

The framework was trained on the RSNA 2020 PE Diffusion Model Split. Input images were clipped to a Hounsfield unit (HU) range of [-1000, 3000] and normalized by subtracting the mean and dividing by the standard deviation, both computed from the training set.

Training proceeded in three stages to improve stability at increasing resolution. First, a low-resolution autoencoder (encoder $E$ and U-Net $\epsilon_\theta$) was trained jointly at $128\times128$ for 5 epochs.
Second, at $256\times256$, $\epsilon_\theta$ was trained for 1 epoch with $E$ frozen, conditioned on semantic vectors extracted from the low-resolution encoder. $E$ was then separately fit for 1 epoch at $256\times256$ to
reproduce the semantic vectors from the low-resolution encoder, with $\epsilon_\theta$ kept fixed. Finally, $E$ and $\epsilon_\theta$ were jointly fine-tuned end-to-end at $256\times256$ for 80 epochs. All stages used the AdamW optimizer with a learning rate of
$1\times10^{-5}$, a batch size of 32, and exponential moving average (EMA) weight updates.
Networks were randomly initialized, and the semantic latent vector dimensionality was fixed at 512.

\subsection*{Synthetic chest CT image generation}

Given a semantic latent vector $z$, image synthesis is performed using the
trained diffusion autoencoder via classifier-free guidance
\cite{ho2022classifierfreediffusionguidance}, starting from Gaussian noise
$x_T \approx \mathcal{N}(0,I)$ and iteratively applying the reverse process
until $t=0$. At each step, the noise prediction is guided by interpolating
between the conditional and unconditional estimates,
\begin{equation}
    \tilde\epsilon_\theta(x_t,t,z) = \epsilon_\theta(x_t,t,\emptyset)
    + w\big(\epsilon_\theta(x_t,t,z) - \epsilon_\theta(x_t,t,\emptyset)\big),
    \label{eq:cfg}
\end{equation}
where $w=6$ is the guidance scale and $\emptyset$ denotes the null (unconditional)
vector. An estimate of the clean image is then recovered and propagated to the
next timestep $t'$ in a DDIM manner \cite{song2022denoisingdiffusionimplicitmodels}:
\begin{equation}
    \hat{x}_0 = f^{-1}\!\left(\frac{x_t - \tilde\epsilon_\theta(x_t,t,z)\,  \sqrt{1-\bar{\alpha}_t}}{\sqrt{\bar{\alpha}_t}},\, t\right),
    \label{eq:x0-estimate}
\end{equation}
\begin{equation}
    x_{t'} = \sqrt{\bar{\alpha}_{t'}}\, f(\hat{x}_0,\, t')
    +  \sqrt{1-\bar{\alpha}_{t'}}\Big(\sqrt{a}\,\tilde\epsilon_\theta(x_t,t,z) + \sqrt{1-a}\,\epsilon'\Big),
    \qquad \epsilon' \sim \mathcal{N}(0,I),
    \label{sample}
\end{equation}
where $a=0.65$ (predicted noise retention) interpolates between deterministic DDIM
sampling ($a=1$) and fully stochastic resampling ($a=0$).

The latent vector $z$ can be obtained in two ways. First, it can be produced directly by the semantic encoder $E$ for a given image, yielding a reconstruction that preserves the same semantic content while introducing stochastic variation.
Second, to generate new images or synthetic datasets, $z$ can instead be sampled from a class-specific distribution estimated separately for each class.
Specifically, for the RSNA 2020 PE and RadFusion datasets, latent vectors were extracted for all images in the respective \textbf{Classifier Training Split} partitions and a KDE (bandwidth $=0.1$) was fit separately to the vectors of PE-positive and PE-negative images.
We used a KDE because, as a non-parametric estimator, it makes no assumption about the number or shape of modes of the class-conditional distributions, which may be multimodal.
To mitigate the risk of regenerating training samples, by accidentally sampling a latent vector $z$ too similar to one derived from a real image, each sampled $z$ was compared to the latent vectors of the real data in the \textbf{Classifier Split} partitions via nearest-neighbor search (L2 distance). Samples with a distance below a threshold were rejected. The threshold was estimated by randomly selecting 3,000 real vectors and computing, for each, the L2 distance to its nearest neighbor in the combined (PE-negative and PE-positive) training pool. The rejection threshold was then set to the \(5^{\text{th}}\) percentile of this empirical real-to-real distance distribution. The distribution of distances is shown in the Supplementary Material.

\subsection*{Fréchet Inception Distance between real and synthetic CT slices}
\label{sec:fid}

Fréchet Inception Distance (FID) \cite{heusel2018ganstrainedtimescaleupdate} was computed with an Inception-v3 backbone pretrained on RadImageNet \cite{Mei2022}. Slices were rescaled from $[-1000, 3000]$\,HU to $[-1,1]$, replicated to 3 channels and resized to $299\times299$ pixels (bilinear interpolation). We computed the FID between the synthetic and real slices of the classifier training split for RSNA 2020 PE and RadFusion (38,000 slices each), and between real RSNA 2020 PE and real RadFusion as a real-vs-real reference. 95\% confidence intervals were obtained by bootstrapping (100 resamples).

To test whether the synthetic-vs-real gap exceeds the variability between two real CT cohorts, each synthetic-vs-real FID was compared with the real-vs-real FID using a two-sided paired bootstrap test on the FID difference (200 resamples). Both FIDs of a comparison share one real dataset as reference (e.g., real RSNA 2020 PE for $\mathrm{FID}(\text{synthetic RSNA}, \text{real RSNA})$ vs.\ $\mathrm{FID}(\text{real RadFusion}, \text{real RSNA})$), which was resampled once per bootstrap iteration and used for both FIDs \cite{efron1993bootstrap}.

\subsection*{Latent-space PE classification}
\label{sec:latent-classification}

To test whether PE status is clearly separable in the diffusion autoencoder's semantic latent space, and whether this separability is preserved after KDE-based resampling, we trained a linear classifier on the 512-dimensional latent vectors $z$ extracted from the \textbf{Classifier Training Split}, separately for each dataset (RSNA 2020 PE and RadFusion).

The used latent classifier is a simple single linear layer with dimension 512, matching the dimension of the latent vector $z$.
For each dataset, the latent vectors from the real images were divided into stratified train/validation/test
splits (60\%;20\%;20\%, preserving the PE-positive/PE-negative ratio in each split).
All latent vectors were standardized per dimension as $\tilde{z}_d = (z_d - \mu_d)/\sigma_d$, where $\mu_d$ and $\sigma_d$ are the mean and standard deviation of dimension $d$ computed on the training split only. These training statistics were then applied, unchanged, to the validation and test splits and to the KDE-sampled latent vectors.
The latent classifier was trained with a binary cross-entropy loss and optimized with Adam (learning rate $10^{-2}$, batch size 2048) for up to 200 epochs. Model selection used early stopping on the validation set, with AUC as the metric (patience of 20 epochs without improvement), and the checkpoint with the highest validation AUC was saved.

The trained latent classifier was then applied to both the held-out real test vectors and the KDE-sampled class-specific latent vectors. The receiver operating characteristic (ROC) curves and AUC were computed for each. 95\% confidence intervals for each AUC were computed analytically using DeLong's method \cite{delong}. Since the real and KDE-sampled vectors form independent samples, their AUCs were compared with a two-sided $z$-test based on the DeLong standard errors.

\subsection*{PE Classifiers}

Classifiers were trained separately for the RSNA 2020 PE and RadFusion datasets, using the ResNet-50 architecture, pretrained on ImageNet \cite{he2015deepresiduallearningimage}. For each dataset, models were trained on randomly sampled \textbf{Real}, \textbf{Mixed}, and \textbf{Synthetic} training sets, as described above, with 10 independently seeded models trained per dataset-variant combination, for a total of 60 classifier models ($2$ datasets $\times$ $3$ variants $\times$ $10$ seeds).
All models were optimized with Adam (default hyperparameters) using a weighted cross-entropy loss. Following inverse-class-frequency weighting, class weights of $[0.25, 0.75]$ (PE-negative, PE-positive) were used for the real and synthetic variants, and uniform weights $[1.0, 1.0]$ for the already class-balanced mixed variant. Training used a base learning rate of $1\times10^{-4}$, decayed each epoch as $lr = lr_0/(\text{epoch}+1)$, and was stopped when validation area under the receiver operating characteristic curve (AUC) failed to improve for 5 consecutive epochs.
Data augmentation included random horizontal and vertical flips (probability 0.5 each) and random crop-and-resize operations (85--100\% of the original image size) with probability 0.7, a random rotation of up to $\pm15^\circ$ was additionally applied, followed by upsampling and a center-crop to avoid empty image corners. Models were validated at the end of every epoch, and the checkpoint achieving the best validation AUC was selected for evaluation.

For inference, image-level prediction probabilities were averaged across the 10-model ensemble. Model performance was primarily assessed using the AUC, with 95\% confidence intervals estimated using DeLong's method \cite{delong}. Statistical significance was evaluated using two-sided tests with $\alpha = 0.05$. Confusion matrices were generated using the threshold that maximized Youden's J statistic (optimizing Sensitivity + Specificity - 1) \cite{Youden1950} on the validation set. With these binary results, accuracy, sensitivity and specificity values were calculated. Wilson score intervals were used to calculate 95\% CIs on those \cite{Wilson1927-wh}. Since all models were evaluated on the same test images, accuracy, sensitivity and specificity of the models trained on synthetic or mixed data were compared with those of the model trained on real data using the exact McNemar test \cite{McNemar1947}, applied to all, PE-positive and PE-negative test images, respectively. P-values of the pairwise DeLong and McNemar tests were adjusted for multiple comparisons within each dataset using the Holm method \cite{Holm1979}; adjusted p-values are reported.

Model explainability was assessed using gradient-weighted class activation mapping (Grad-CAM) saliency maps \cite{Selvaraju_2019}, computed using the Captum library \cite{captum2024}.

\subsection*{Memorization analysis}
To test whether the framework produces genuinely new samples rather than copies of real images, we compared each of the 38,000 synthetic RSNA 2020 PE slices against all slices of the two real sets involved in its generation: the \textbf{Diffusion Model Split}, on which the diffusion autoencoder was trained, and the \textbf{Classifier Training Split}, to whose latent vectors the KDE was fitted (200,000 slices each). All slices were clipped to $[-1000, 3000]$\,HU and resized to $256\times256$ pixels before comparison.

Similarity was measured with two complementary metrics. The first, self-supervised copy-detection (SSCD) similarity, is the cosine similarity between embeddings from a self-supervised descriptor for image copy detection \cite{pizzi2022selfsuperviseddescriptorimagecopy}. For this metric, each slice was scaled to $[0,1]$ within three HU windows -- lung ($-1350$ to $150$\,HU), mediastinal ($-160$ to $240$\,HU), and bone ($-450$ to $1050$\,HU) -- stacked as a three-channel image, resized to $288\times288$ pixels, and normalized with the ImageNet mean and standard deviation. The second, pixel-intensity correlation, is the Pearson correlation between two slices, a simple pixel-based measure previously used to detect memorization in medical diffusion models \cite{Akbar2025}. For this metric, slices were clipped to $[-1000, 1000]$\,HU and downsampled to $64\times64$ pixels.
In a validation on simulated copies of real CT slices (Supplementary Material), pixel-intensity correlation reliably detected copies that preserve the pixel layout, whereas SSCD also detected geometrically transformed copies. For each synthetic slice, the most similar real slice was identified separately under each metric.

\subsection*{Ablation experiments}
\textbf{Conditional diffusion baseline.} Instead of training the diffusion autoencoder on unlabeled data and modeling the class-conditional distribution in its latent space, a conditional diffusion model can be trained directly on labeled data. As a baseline, we trained such a model on the \textbf{Classifier Training Split} (200,000 slices, 9,891 PE-positive), using the same U-Net architecture, blurring diffusion process and training procedure as the diffusion autoencoder: 5 epochs at $128\times128$, followed by 81 epochs at $256\times256$, with AdamW (learning rate $1\times10^{-5}$), batch size 32 and an exponential moving average (EMA) of the weights. The only difference was the conditioning: instead of the semantic latent vector $z$, the U-Net received one of two fixed random vectors (PE-negative or PE-positive), drawn once from a normal distribution with the dimension and scale of $z$. These vectors serve only as class labels, whose meaning the network learns during training. To enable classifier-free guidance, 10\% of the condition vectors were replaced by the null vector $\emptyset$ during training. Unlike the diffusion autoencoder, which is trained on separate, unlabeled data and uses the labeled slices only to fit the latent distribution, this model thus learns the image distribution directly from the labeled slices the classifiers are trained on.
Using the EMA weights of the final epoch and the same sampler, we generated a synthetic dataset of the same size and class ratio as the main synthetic dataset (28,500 PE-negative, 9,500 PE-positive slices). Because the sampling settings of the diffusion autoencoder ($w=6$, $a=0.65$) produced unstable images with this model, we visually inspected a grid of settings ($w$: 2--8, $a$: 0.35--0.85; Supplementary Material) and chose $w=5$ and $a=0.45$, which consistently produced realistic images while staying closest to the settings of the diffusion autoencoder. PE classifiers were then trained on this synthetic dataset as described above.

\textbf{Latent distribution.} To assess the choice of the KDE, we replaced it with two simpler parametric distributions, fitted per class to the same latent vectors: (i) a multivariate Gaussian with full covariance matrix (class mean and covariance; 132k parameters per class), and (ii) a diagonal Gaussian, which models each latent dimension independently (mean and variance per dimension; 1,024 parameters per class). Unlike the KDE, which places a narrow Gaussian kernel on every real latent vector, both describe each class by a single global distribution. All other steps were identical to the main pipeline: the same diffusion autoencoder, class counts (28,500 PE-negative, 9,500 PE-positive), latent-distance rejection, sampler settings, and PE classifier training.
For the conditional diffusion baseline and both latent distributions, the resulting ensemble AUC was compared to that of the proposed DiffAE~+~KDE approach using DeLong's test, with $p$-values adjusted for multiple comparisons using the Holm method \cite{Holm1979} (Table~\ref{tab:method_comparison}).

\textbf{Architecture.} We additionally ablated two components of the diffusion autoencoder, blurring diffusion and the gated residual block, by comparing reconstruction quality during the first training epochs (Supplementary Material).

\subsection*{Implementation details and Code availability}
The following software versions were used for this study: Python 3.9; PyTorch 2.4.1; torchvision 0.19.1; numpy 1.24.4; matplotlib 3.7.5; pydicom 2.4.4; tqdm 4.66.5. All models were trained on a single NVIDIA A100 (80\,GB VRAM). The source code is publicly available at \url{https://github.com/J-3TO/SynthCT-DiffAE}.

\section*{Results}\label{results}
We trained the diffusion autoencoder only on the RSNA 2020 PE dataset Diffusion Model Split, as described in the Methods section. All subsequent data generation and analyses were performed independently for the RSNA 2020 PE and RadFusion datasets, without any mixing of data between databases.

\subsection*{Generated Images}
\begin{figure}[h]
    \centering
    \includegraphics[width=1.0\linewidth]{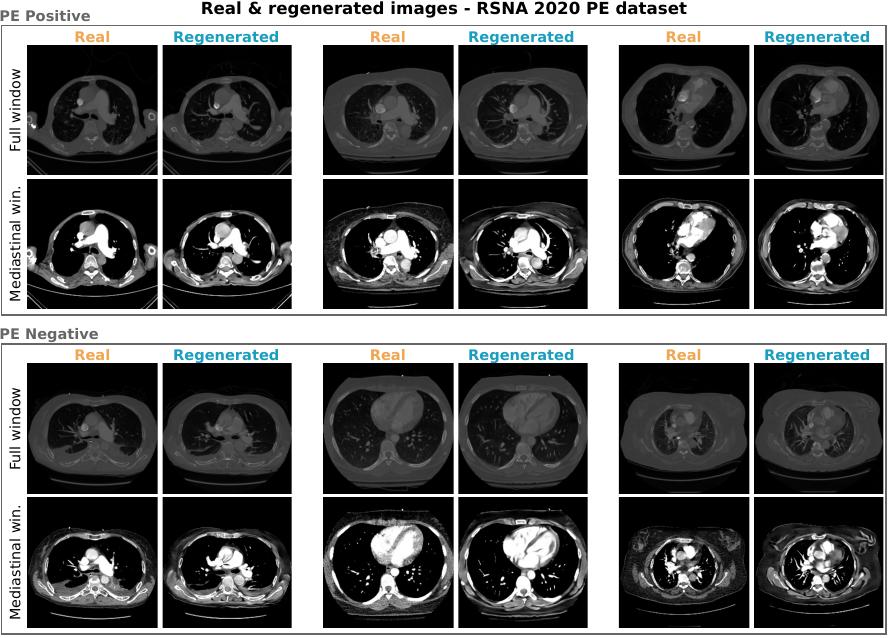}
    \caption{Original PE-positive and PE-negative slices from the RSNA Test Split alongside synthetic images generated by conditioning the generative model on the latent vectors $z$ derived from the corresponding real images (shown adjacent to each pair). Each image is displayed in both the full-range window (L: 1,000 HU; W: 4,000 HU) and the mediastinal window (L: 50 HU; W: 380 HU).}
    \label{fig:ReGen_RSNA}
\end{figure}

\begin{figure}[h]
    \centering
    \includegraphics[width=1.0\linewidth]{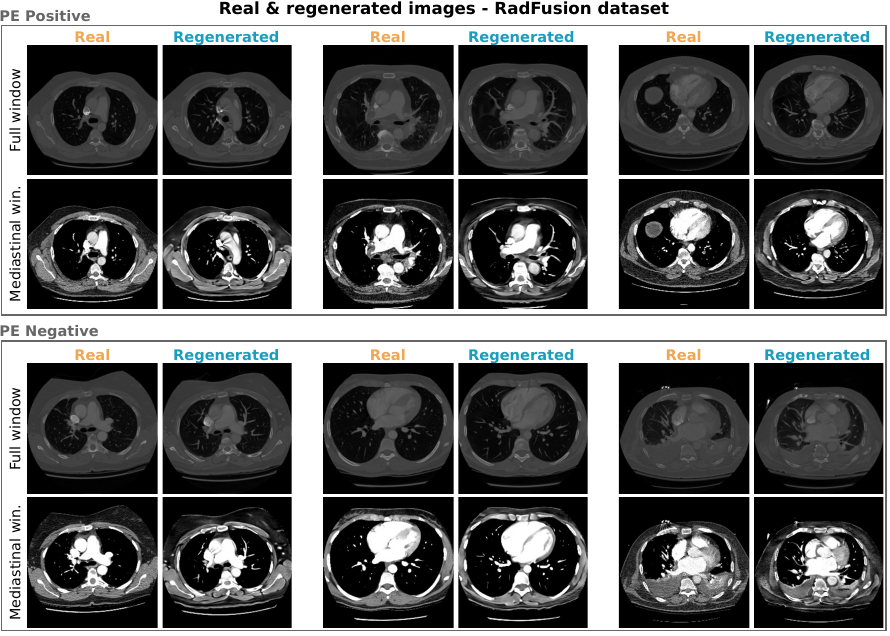}
    \caption{Original PE-positive and PE-negative slices from the RadFusion Test Split alongside synthetic images generated by conditioning the generative model on the latent vectors $z$ derived from the corresponding real images (shown adjacent to each pair). Each image is displayed in both the full-range window (L: 1,000 HU; W: 4,000 HU) and the mediastinal window (L: 50 HU; W: 380 HU).}
    \label{fig:ReGen_RadFusion}
\end{figure}

Figures~\ref{fig:ReGen_RSNA} and \ref{fig:ReGen_RadFusion} show representative examples of real images and their regenerated counterparts from the RSNA 2020 PE dataset and the RadFusion dataset, obtained by conditioning the generative model on the latent vector $z$ extracted from each real image.
In both figures, PE-positive and PE-negative images with their regenerated counterparts are presented. The reconstructed images closely resemble the originals while exhibiting a slightly smoother appearance.

\begin{figure}[h]
    \centering
    \includegraphics[width=1.0\linewidth]{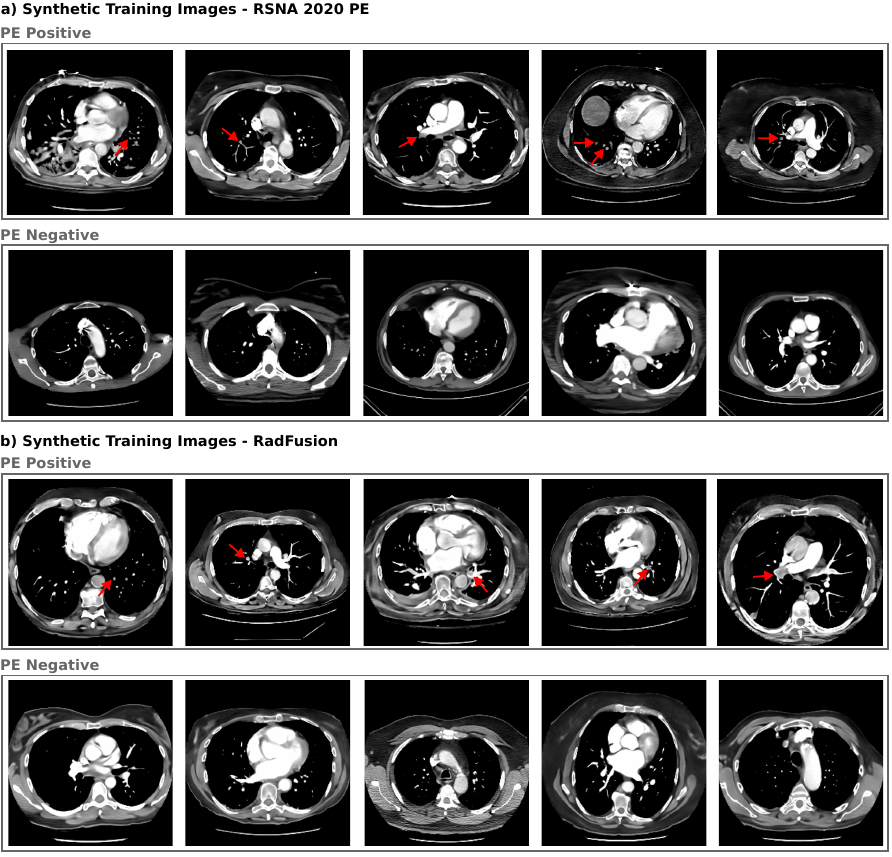}
    \caption{Representative synthetic CT images generated by sampling new semantic latent vectors $z$ from per-class Gaussian kernel density estimates (KDEs) fit on the real PE-positive and PE-negative latent vectors of each dataset's classifier training split. The sampled vectors were then decoded with the generative model. a) PE-positive (top) and PE-negative (bottom) examples from latent vectors sampled from the RSNA 2020 PE training split. b) The same for the RadFusion training split. Displayed in the mediastinal window (W: 380 HU; L: 50 HU). The red arrows indicate visible emboli.}
    \label{fig:KDE_samples}
\end{figure}

Figure~\ref{fig:KDE_samples} illustrates fully synthetic image generation using newly sampled semantic latent vectors with distance filtering. Figure \ref{fig:KDE_samples}~(a) shows PE-positive and PE-negative images generated from latent vectors sampled from KDEs fitted to the RSNA 2020 PE Classifier Training Split, while Figure \ref{fig:KDE_samples}~(b) presents samples generated from KDEs fitted to the RadFusion Classifier Training split. PE positive regions (red arrows) are clearly visible in the presented images generated from the latent vectors sampled from the PE positive latent distributions from the RSNA 2020 PE and the RadFusion dataset.

\begin{figure}[h]
	\centering
	\includegraphics[width=1.0\textwidth]{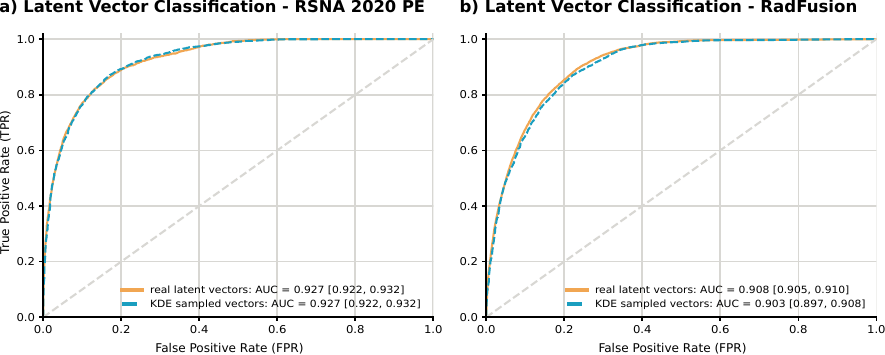}
	\caption{Performance of the linear classifier trained on real latent vectors, evaluated on a held-out test set of 40,000 real latent vectors, and on latent vectors resampled from the fitted class-conditional KDE distributions. \textbf{a)} Receiver operating characteristic (ROC) curves for latent vectors derived from the RSNA 2020 PE test dataset. \textbf{b)} ROC curves for latent vectors derived from the external RadFusion dataset.}
    \label{fig:LatentClassification}
\end{figure}

Figure~\ref{fig:LatentClassification} summarizes the performance of the linear classifier trained directly on latent vectors from the RSNA 2020 PE and RadFusion dataset, respectively. For the RSNA 2020 PE dataset vectors, the classifier achieved an AUC of 0.927 (95\% CI: 0.922--0.932) on both real and KDE-sampled latent vectors ($p = 0.913$), indicating that the sampled latent representations preserve the strong class separation present in the original latent space. Similar results were observed for RadFusion, although with slightly lower performance: the classifier achieved an AUC of 0.908 (95\% CI: 0.905--0.910) on real latent vectors and 0.903 (95\% CI: 0.897--0.908) on KDE-sampled vectors ($p = 0.113$). For neither dataset did the AUC on KDE-sampled vectors differ significantly from that on real vectors. Overall, these findings suggest that PE-positive and PE-negative samples remain well separated in the latent space and that this structure is largely preserved by KDE-based sampling.

The synthetic-vs-real images yielded an FID of 0.055 (95\% CI [0.054, 0.056]) for RSNA 2020 PE and 0.076 (95\% CI [0.074, 0.079]) for RadFusion, both significantly higher than the real-vs-real baseline of 0.009 (95\% CI [0.009, 0.010]), with $p < 0.01$ for both (the smallest $p$-value attainable with 200 bootstrap resamples). This indicates a distributional gap between the synthetic and real slices that goes beyond the variation between two real CT cohorts.

\subsection*{Classifier performance}
To assess the diagnostic utility of the generated data, classifiers were either trained on real, synthetic, or mixed datasets, with the latter consisting of balanced PE positive and negative cases, as described in the Methods section, and evaluated for PE detection on real test data. Performance was averaged across ten independently trained models per experimental configuration, with the training and validation sets resampled for each run.

\begin{figure}[h]
	\centering
	\includegraphics[width=1.0\textwidth]{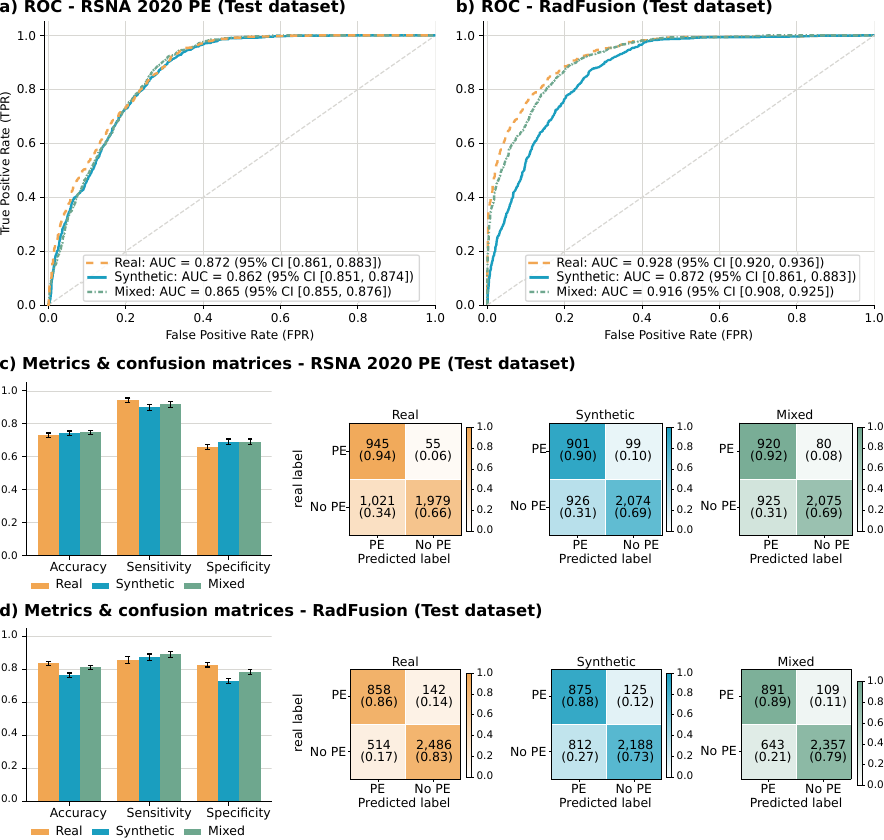}
	\caption{Results of the classification models trained on either real, synthetic, or mixed data, averaged over 10 independently-trained runs. \textbf{a)} Receiver Operating Characteristic (ROC) curve on the RSNA 2020 PE test dataset. \textbf{b)} ROC curve on the RadFusion test dataset. \textbf{c)} Accuracy, sensitivity, and specificity and the corresponding confusion matrices on the RSNA 2020 PE test and \textbf{d)} RadFusion test datasets. Threshold values were determined by maximizing Youden’s J statistic, respectively.}
	\label{fig:Classification_Results}
\end{figure}

Results for the RSNA 2020 PE and RadFusion test splits are summarized in Figure~\ref{fig:Classification_Results} and Table~\ref{tab:classification_performance}. Figures~\ref{fig:Classification_Results}(a) and \ref{fig:Classification_Results}(b) show the ROC curves for RSNA 2020 PE and RadFusion, respectively.
Accuracy, sensitivity, specificity, and the corresponding confusion matrices are shown in Figures~\ref{fig:Classification_Results}(c) and \ref{fig:Classification_Results}(d).

\begin{table}[h]
\centering
\scriptsize
\setlength{\tabcolsep}{3pt}
\caption{Classifier performance (mean, 95\% CI) for the real, synthetic, and mixed training variants, by dataset. $p$: comparison to the real-data classifier (DeLong's test for AUC; McNemar's test for accuracy, sensitivity, and specificity), Holm-adjusted for multiple comparisons.}
\label{tab:classification_performance}
\resizebox{\linewidth}{!}{%
\begin{tabular}{ll cc cc cc cc}
\toprule
& & \multicolumn{2}{c}{AUC} & \multicolumn{2}{c}{Accuracy} & \multicolumn{2}{c}{Sensitivity} & \multicolumn{2}{c}{Specificity} \\
\cmidrule(lr){3-4} \cmidrule(lr){5-6} \cmidrule(lr){7-8} \cmidrule(lr){9-10}
Dataset & Variant & Value (95\% CI) & $p$ & Value (95\% CI) & $p$ & Value (95\% CI) & $p$ & Value (95\% CI) & $p$ \\
\midrule
\multirow{3}{*}{RSNA 2020 PE}
 & Real      & 0.872 (0.861--0.883) & --     & 0.731 (0.717--0.745) & --     & 0.945 (0.929--0.958) & --     & 0.660 (0.643--0.676) & --     \\
 & Synthetic & 0.862 (0.851--0.874) & 0.007  & 0.744 (0.730--0.757) & 0.005  & 0.901 (0.881--0.918) & $<0.001$ & 0.691 (0.675--0.708) & $<0.001$ \\
 & Mixed     & 0.865 (0.855--0.876) & 0.004  & 0.749 (0.735--0.762) & $<0.001$ & 0.920 (0.902--0.935) & $<0.001$ & 0.692 (0.675--0.708) & $<0.001$ \\
\midrule
\multirow{3}{*}{RadFusion}
 & Real      & 0.928 (0.920--0.936) & --     & 0.836 (0.824--0.847) & --     & 0.858 (0.835--0.878) & --     & 0.829 (0.815--0.842) & --     \\
 & Synthetic & 0.872 (0.861--0.883) & $<0.001$ & 0.766 (0.752--0.779) & $<0.001$ & 0.875 (0.853--0.894) & 0.196  & 0.729 (0.713--0.745) & $<0.001$ \\
 & Mixed     & 0.916 (0.908--0.925) & $<0.001$ & 0.812 (0.800--0.824) & $<0.001$ & 0.891 (0.870--0.909) & 0.001  & 0.786 (0.771--0.800) & $<0.001$ \\
\bottomrule
\end{tabular}%
}
\end{table}

For the RSNA 2020 PE dataset, classifiers trained on real, synthetic, and mixed data showed closely comparable performance across all metrics; AUC reductions relative to real-data training were statistically significant but small for both the synthetic- and mixed-data classifiers (Table~\ref{tab:classification_performance}).

For the RadFusion dataset, classifiers trained on synthetic data performed worse than those trained on real data, with classifiers trained on the mixed dataset falling in between the two; AUC reductions relative to real-data training were statistically significant for both the synthetic- and mixed-data classifiers (Table~\ref{tab:classification_performance}).

\begin{figure}[h]
	\centering
	\includegraphics[width=1.0\textwidth]{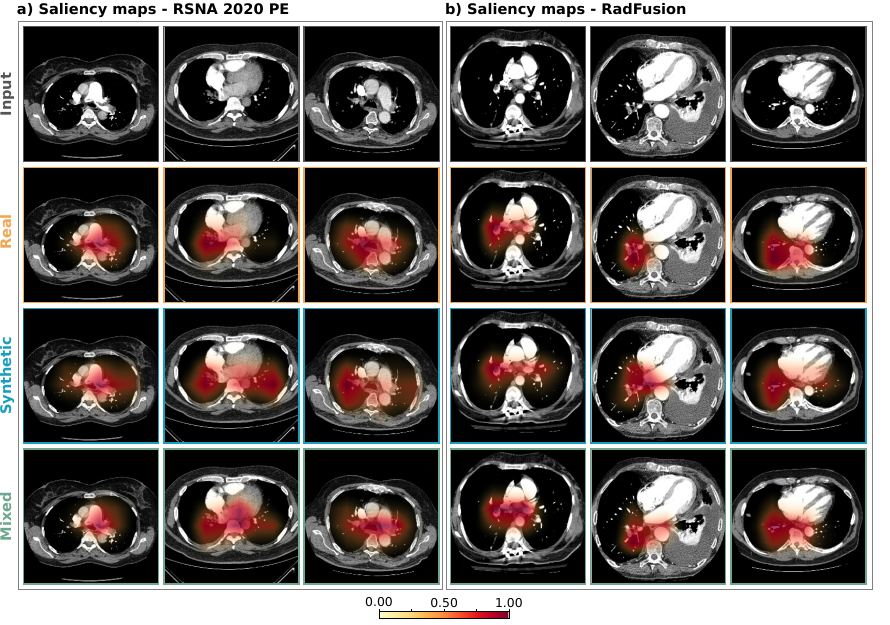}
 	\caption{Grad-CAM saliency maps, averaged across 10 independently-trained runs per variant, for classifiers trained on either real, synthetic, or a balanced mix of real and synthetic images. Each column shows the input image (mediastinal window, WW=380/WL=50 HU) alongside the corresponding overlay for each variant. Example images from \textbf{a)} the RSNA 2020 PE held-out test set and \textbf{b)} the RadFusion held-out test set.}
	\label{fig:sal_maps}
\end{figure}

To further examine differences in classifier behavior and assess whether the models, particularly those trained on synthetic data, relied on clinically relevant image features rather than potential shortcut signals, we compared Grad-CAM saliency maps across the different training configurations. Representative PE-positive test images and their corresponding saliency maps are shown in Fig.~\ref{fig:sal_maps}. Qualitatively, the highlighted regions generally corresponded to areas containing pulmonary emboli and appeared broadly consistent across the real, synthetic, and mixed classifier variants.

\subsection*{Memorization analysis}

\begin{figure}[h]
	\centering
	\includegraphics[width=1.0\textwidth]{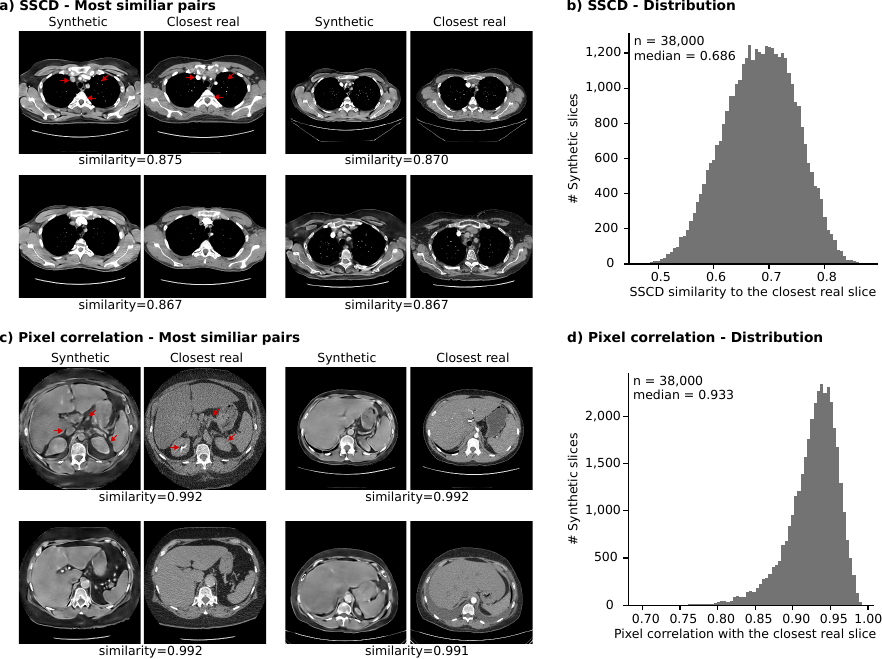}
    \caption{\textbf{Nearest-neighbor check for training-data memorization.}
    For each of the 38,000 synthetic RSNA 2020 PE slices, the most similar real slice in the Diffusion Model Split and the Classifier Training Split (200,000 slices each) was identified using self-supervised copy-detection (SSCD) similarity (\textbf{a,~b}) and pixel-intensity correlation (\textbf{c,~d}).
    \textbf{a,~c)} The four synthetic slices most similar to a real slice, each shown next to its closest real match (mediastinal window, $-160$ to $240$\,HU). Red arrows mark differences between the synthetic and real slice in the most similar pair.
    \textbf{b,~d)} Distribution of the similarity of all 38,000 synthetic slices to their closest real match (medians 0.686 and 0.933, respectively).}

    \label{fig:memorization}
\end{figure}

To test whether the framework reproduces real training images, we searched both real training sets for each synthetic slice's closest match (Fig.~\ref{fig:memorization}). Even the closest synthetic--real pairs under either metric differed visibly: they showed comparable anatomy at a similar slice position, but different vessel and lung shapes and surrounding soft tissue (Fig.~\ref{fig:memorization}a,~c). Most synthetic slices were considerably less similar to their closest real match than these pairs (median similarity 0.686 for SSCD and 0.933 for pixel-intensity correlation; Fig.~\ref{fig:memorization}b,~d). Thus, neither metric detected direct copies of real training slices.

\subsection*{Ablation experiments}

\begin{table}[h]
\centering
\scriptsize
\setlength{\tabcolsep}{4pt}
\caption{Comparison of latent-sampling strategies and a conditional diffusion baseline for the RSNA 2020 PE classifier, by ensemble AUC. $p$: comparison to the proposed DiffAE~+~KDE approach (DeLong's test, Holm-adjusted).}
\label{tab:method_comparison}
\begin{tabular}{lcc}
\toprule
Training data & AUC (95\% CI) & $p$ (DeLong) \\
\midrule
Real data (reference) & 0.872 (0.861--0.883) & -- \\
\midrule
DiffAE + KDE (proposed) & 0.862 (0.851--0.874) & -- \\
Conditional diffusion (CFG) & 0.777 (0.761--0.792) & $<0.001$ \\
DiffAE + Gaussian & 0.865 (0.854--0.876) & 0.522 \\
DiffAE + diagonal Gaussian & 0.843 (0.830--0.856) & $<0.001$ \\
\bottomrule
\end{tabular}
\end{table}

Classifiers trained on synthetic data from the proposed DiffAE~+~KDE approach reached an AUC close to the real-data reference (0.862 vs.\ 0.872; Table~\ref{tab:method_comparison}). The conditional diffusion baseline performed markedly worse (AUC 0.777, $p<0.001$), supporting the benefit of the semantic latent space over direct class-conditional generation. Replacing KDE with a full-covariance Gaussian fit gave comparable performance to the proposed approach (AUC 0.865, $p=0.522$), whereas a diagonal-covariance Gaussian performed significantly worse (AUC 0.843, $p<0.001$), indicating that capturing correlations between latent dimensions, but not necessarily their full non-parametric distribution, is important for generating diagnostically useful synthetic data. Ablating either blurring diffusion or the gated residual block increased the reconstruction error after five training epochs (MAE 244 and 195\,HU, respectively, vs.\ 160\,HU for the proposed model; both $p<0.001$; Supplementary Material).
\clearpage

\section*{Discussion}

In this study, we utilized and evaluated a diffusion autoencoder framework for generating diagnostically meaningful synthetic chest CT images from proprietary datasets, incorporating a latent-space filtering strategy to reduce privacy risk. The approach combines a generative model with class-conditional latent-space density estimation, enabling the generation of labeled synthetic images without retraining the generative model for new downstream tasks.

The diffusion autoencoder learned a semantic latent representation that encodes diagnostically relevant information despite being trained in a fully PE label-free manner.
This is supported by the latent-space classification experiments, in which a simple linear classifier achieved high AUC on both real and KDE-sampled latent vectors. The similarity in performance between the two suggests that the class-specific structure of the latent space is largely preserved during KDE-based density estimation and resampling.
Images generated from real latent vectors preserved anatomical structures while introducing stochastic variation, indicating that the latent representation captures sufficient information to guide realistic image synthesis.
Synthetic images sampled from the PE-positive latent distribution also exhibited visually recognizable embolic findings, further supporting that class-relevant information is encoded in, and recoverable from, the latent space.

The downstream classification experiments demonstrate the practical utility of the generated data. For the RSNA 2020 PE dataset, classifiers trained exclusively on synthetic images performed comparably to classifiers trained on real images, indicating that the synthetic data retain sufficient diagnostic information to support training of clinically relevant predictive models. Results on the RadFusion dataset provide an additional test of generalizability: although the diffusion autoencoder was trained exclusively on RSNA 2020 PE data, synthetic images generated from RadFusion latent vectors still enabled training of accurate PE classifiers, albeit with a larger real-synthetic performance gap than for RSNA 2020 PE. This larger gap may reflect that the generative model was trained exclusively on RSNA data, which could constrain its ability to reproduce RadFusion-specific image features. Differences between the two datasets in acquisition protocol, patient population, or annotation practice may also contribute. These remain plausible interpretations rather than experimentally established causes. More broadly, this suggests that generalizability may be reduced when applying the framework to datasets that differ substantially from the diffusion model's training distribution, motivating evaluation across more diverse cohorts in future work.

Despite this gap, the strong performance of classifiers trained on synthetic RadFusion data indicates that the framework can still transfer diagnostically relevant information across datasets, not merely reproduce the training distribution it was fit on. Grad-CAM analyses further suggested that classifiers trained on real, synthetic, and mixed data focused on similar image regions, providing preliminary qualitative evidence that synthetic-data models learn clinically meaningful features rather than shortcut signals.

We also tested whether the framework could serve as a data-balancing augmentation strategy. In the configurations tested, classifiers trained on a mix of real and synthetic data did not outperform those trained on real data alone, suggesting that under the class imbalance and data volumes studied here, added synthetic images did not provide complementary information beyond what real data already captured.

In the ablation experiments, the conditional diffusion baseline, trained directly on the labeled slices, performed markedly worse than the proposed approach, supporting the benefit of conditioning generation on image-specific semantic latent vectors rather than on class labels alone. Replacing the KDE with a full-covariance Gaussian gave comparable performance, whereas a diagonal Gaussian performed significantly worse, suggesting that for image-level PE classification, modeling correlations between latent dimensions is essential, while a single mode per class suffices. We nevertheless used the KDE, as it does not assume a particular distribution shape and may therefore transfer better to more heterogeneous classes. Conversely, a Gaussian may be preferable when samples should not depend on individual training vectors, as each KDE sample is a perturbed version of one real vector. Since the framework is agnostic to the density estimator, this choice can be adapted to the task and requirements.

From a privacy perspective, the framework introduces a practical safeguard against unintended regeneration of real images. By rejecting sampled latent vectors that are excessively similar to those from the real distribution, a simple regeneration of real images is impeded. Consistent with this, our memorization analysis found no direct copies among the 38,000 synthetic RSNA 2020 PE slices, with even the closest real--synthetic matches differing visibly. This result should be treated as a lower bound on privacy protection for the samples and similarity metrics tested, rather than as a guarantee that memorization cannot occur more broadly.

Several limitations should be acknowledged. First, FID analysis revealed a measurable distributional gap between synthetic and real images, indicating that the synthetic images do not fully capture the variability of the original data and that further gains in generative fidelity are likely possible. More powerful generative models, such as latent diffusion models or larger foundation-scale medical image generators, could improve this aspect in future work.
Second, the framework still requires a sufficiently large labeled dataset to reliably estimate class-specific latent distributions, limiting its applicability to very rare diseases with few labeled examples. Third, while the latent-distance rejection strategy reduces the likelihood of reproducing individual training samples, it provides no formal privacy guarantee and cannot fully exclude the generation of images that closely resemble specific patients. Complementary privacy-preserving techniques and more rigorous privacy-auditing methods should be explored in future work.

Lastly, PE-related features were not consistently reproduced in the synthetic images and were occasionally degraded even in the direct regeneration of real images. This likely reflects that PE signs are subtle and typically visible only in a comparatively narrow mediastinal window, whereas the diffusion model was trained in a label-free manner on the full HU range and therefore had no explicit incentive to preserve these localized features during encoding and decoding.
Neither representative visual examples nor latent-space classification can guarantee that an individual synthetic image is a diagnostically valid representation of pulmonary embolism: latent-space classification shows that PE-related information is statistically separable across many latent vectors, but does not warrant that every single decoded image is faithful to that information. Likewise, the depicted representative synthetic images show that the pipeline can produce recognizable embolic findings, but do not guarantee that every synthetic image does. This reflects the stochastic nature of the generation process, in which decoding and latent-sampling errors can occur. Imperfect or inconsistent PE labeling in the underlying clinical data may further propagate into the model.

Despite these limitations, both the latent-space and image-level classification results indicate good separability between PE-positive and PE-negative cases, suggesting that these imperfections did not prevent the synthetic dataset overall from remaining diagnostically useful.

In conclusion, we show that diffusion autoencoders combined with class-conditional latent-space density estimation can generate synthetic CT images that retain diagnostically meaningful information. These images support training of competitive PE classifiers and offer a practical mechanism for creating privacy-aware synthetic datasets from proprietary clinical data that can be shared and used more freely than the original data. A systematic, radiologist-led review of generated images prior to data sharing could provide a direct safeguard against sharing mislabeled synthetic data and is a promising direction for future work. These findings suggest that latent-space generative modeling may be a promising tool for enabling data sharing, collaborative research, and AI development in settings where access to real patient data is restricted. Future studies should evaluate the clinical realism of generated images and investigate their utility across a broader range of downstream clinical and machine learning tasks.

\section*{Statements \& Declarations}
\subsection*{Funding}
We acknowledge financial support through the European Research Council (ERC Synergy Grant SmartX, SyG 101167328), and the Free State of Bavaria under the Excellence Strategy of the Federal Government and the Länder, as well as by the Technical University of Munich -- Institute for Advanced Study.
\subsection*{Competing Interests}
The authors have no relevant financial or non-financial interests to disclose.

\subsection*{Author Contributions}
All authors contributed to the conception and design of the study. Johannes B. Thalhammer and Kacper Ogórek share first authorship: the diffusion model was developed and implemented by Kacper Ogórek as part of his Master's thesis under the supervision of Johannes B. Thalhammer, who conducted the downstream analyses and led the preparation of the manuscript. Florian Schaff provided general scientific and technical guidance for the study and contributed to manuscript editing, and Daniela Pfeiffer provided radiological guidance for evaluating the generated images. All authors commented on previous versions of the manuscript and read and approved the final version.

\subsection*{Ethics approval}
This study is a secondary, retrospective analysis of two publicly available, fully de-identified datasets (the RSNA 2020 PE dataset and the RadFusion dataset); no new patient data were collected.

\subsection*{Consent to participate}
This study used publicly available, de-identified datasets for which informed consent (or an applicable waiver) was obtained by the original data providers (RSNA and the Stanford Center for Artificial Intelligence in Medicine and Imaging) prior to public release. No additional consent to participate was required for this secondary analysis.

\subsection*{Consent to publish}
Not applicable. All CT images shown in this manuscript are de-identified and do not contain any information that could be used to identify an individual patient.

\subsection*{Data Availability}
The RSNA 2020 PE dataset and the RadFusion dataset analyzed in this study are publicly available at \url{https://www.kaggle.com/competitions/rsna-str-pulmonary-embolism-detection} and \url{https://aimi.stanford.edu/datasets/radfusion-multimodal-pulmonary-embolism-dataset}, respectively.

\section*{Acknowledgments}
We thank Jason Alan Fries from Stanford University for assistance with obtaining the RadFusion dataset.

\bibliographystyle{vancouver}
\bibliography{bibliography}

\clearpage
\setcounter{figure}{0}
\setcounter{table}{0}
\setcounter{equation}{0}
\renewcommand{\thefigure}{S\arabic{figure}}
\renewcommand{\thetable}{S\arabic{table}}
\renewcommand{\theequation}{S\arabic{equation}}
\renewcommand{\theHfigure}{S\arabic{figure}}
\renewcommand{\theHtable}{S\arabic{table}}
\renewcommand{\theHequation}{S\arabic{equation}}

\section*{Supplementary Material}

\subsection*{Diffusion Schedule}

The forward process is indexed by a discrete timestep $\bar{t} \in \{a_0, \dots,
a_1\}$,  $\bar{t} = a_1$ corresponds to an essentially
uncorrupted acquisition, while $\bar{t} = a_0$ corresponds to the maximally corrupted acquisition. We further define the normalized timestep:
\begin{equation}
    t = \frac{a_1 - \bar{t}}{a_1 - a_0} \in [0, 1].
    \label{eq:t-bar}
\end{equation}
In our experiments we use $a_0=3$, $a_1=180$, resulting in 178 timesteps. \\
The signal ratio schedule is defined by
\begin{equation}
    \sqrt{\bar{\alpha}_t} = \cos\!\Big(\arccos(0.95) + t \big(\arccos(0.2)-\arccos(0.95)\big)\Big),
    \label{eq:sr}
\end{equation}
with 0.2, and 0.95 as the boundary values of the $\sqrt{\bar{\alpha}_t}$.

The blurring operator $f$ (and its inverse $f^{-1}$) model the progressive loss of
high-frequency image content by applying a low-pass filter in the Fourier domain,
\begin{equation}
    f(x, t) = \mathcal{F}^{-1}\big[H(\omega, t) \odot \mathcal{F}[x]\big],
    \qquad
    f^{-1}(x, t) = \mathcal{F}^{-1}\!\left[\frac{\mathcal{F}[x]}{H(\omega, t)}\right],
    \label{eq:blur-deblur}
\end{equation}
where $\mathcal{F}$, $\mathcal{F}^{-1}$ denote the 2D discrete Fourier transform and its
inverse, and $\odot$ is elementwise multiplication. The transfer function
\begin{equation}
    H(\omega,t) = (1-\lambda)\, e^{-\lVert\omega\rVert^2\, \tau(t)} + \lambda,
    \qquad
    \tau(t) = \frac{1}{2}\left[\left(\frac{0.1}{0.115 - 0.1\,t}\right)^{\!2} - 0.756\right]^{2},
    \label{eq:freq-mask}
\end{equation}
is the frequency response of an isotropic heat-diffusion (Gaussian blur) kernel with
dissipation time $\tau(t)$: attenuation grows with spatial frequency $\lVert\omega\rVert$
and with $t$, so higher diffusion timesteps correspond to a more aggressively blurred
image, mirroring how coarser, less informative image evidence loses fine detail. The
floor $\lambda=0.001$ keeps $H$ bounded away from zero, which is what makes $f^{-1}$
(plain division by $H$) numerically well-defined rather than blowing up at high
frequencies. $\tau(t)$ itself is not derived from the heat equation directly but is an
heuristically calibrated reparameterization of the linear diffusion time $t\in[0,1]$, chosen
so that the resulting blur strength increases at a rate that matches the corruption
schedule used elsewhere in the model. This approach is inspired by \cite{hoogeboom2024blurringdiffusionmodels}.

\newpage

\subsection*{Model Architectures}

\begin{figure}[h!]
    \centering
    \includegraphics[width=1.0\linewidth]{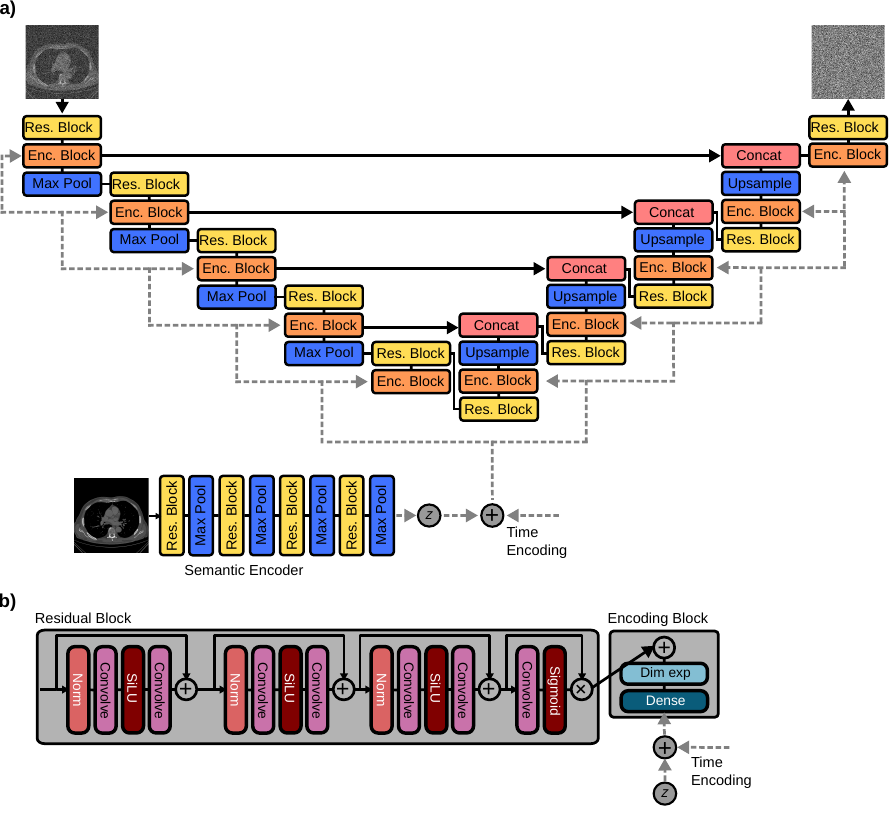}
    \caption{\textbf{A} Detailed architecture of the used U-Net backbone and semantic encoder. \textbf{B} Structure of the Residual and the Encoding Block.}
    \label{fig:sup_architecture}
\end{figure}

\newpage
\subsection*{Hyperparameter Dependence}

\begin{figure}[h!]
    \centering
    \includegraphics[width=1.0\linewidth]{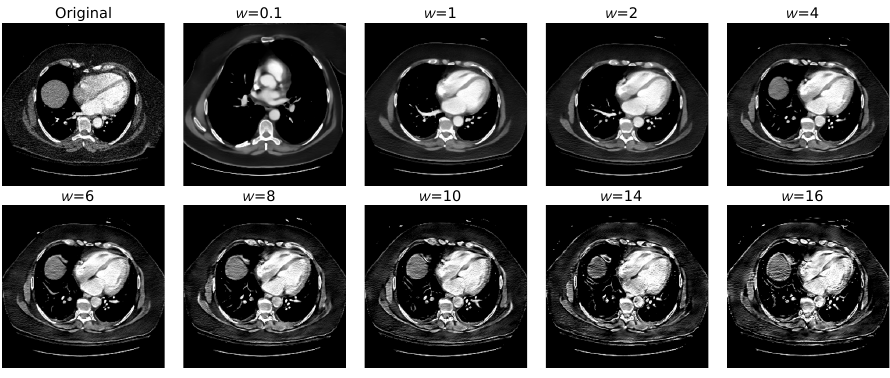}
    \caption{Effect of the classifier-free guidance scale $w$ on the image generation process.}
    \label{fig:sup_CFG}
\end{figure}

\begin{figure}[h!]
    \centering
    \includegraphics[width=1.0\linewidth]{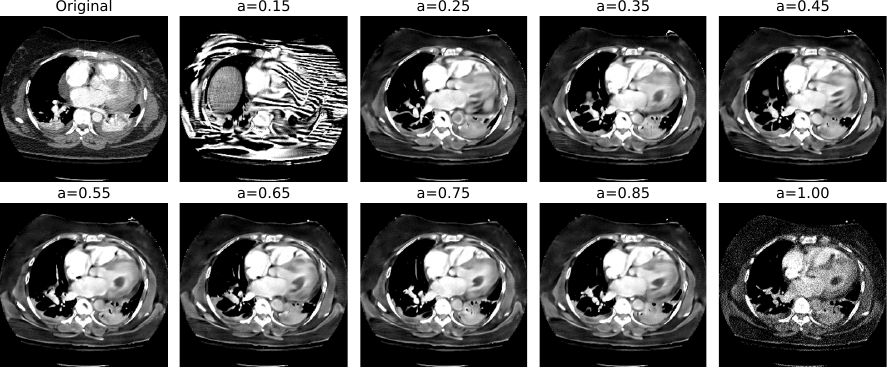}
    \caption{Effect of the predicted noise retention parameter $a$ on stochastic variation during the image generation process.}
    \label{fig:sup_noiseRetention}
\end{figure}

\newpage
\subsection*{Latent vector distances}

\begin{figure}[h!]
    \centering
    \includegraphics[width=.7\linewidth]{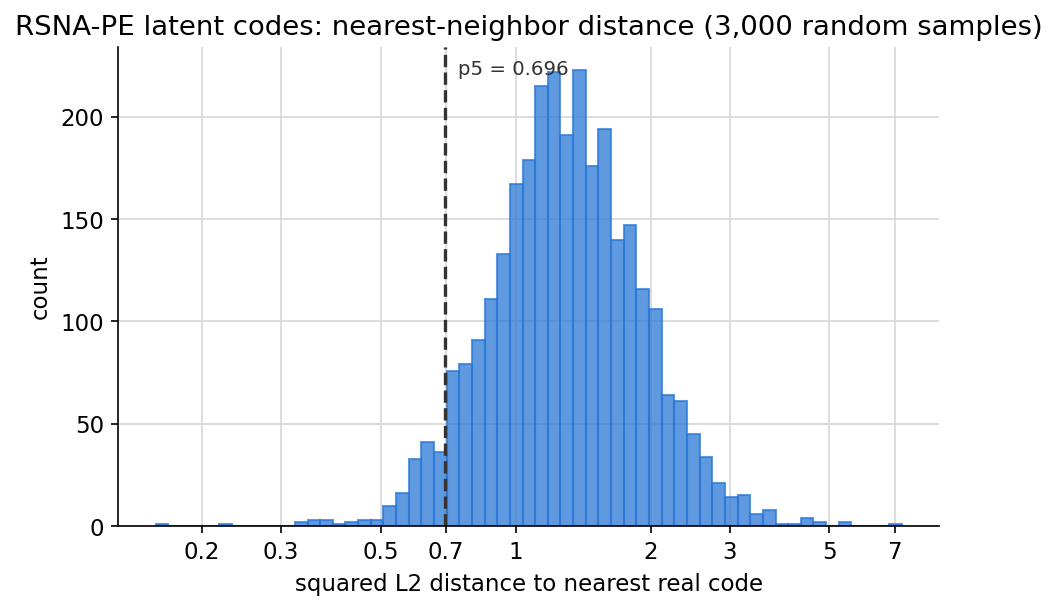}
    \caption{Plot of the nearest neighbor (L2) distances of randomly sampled latent vectors derived from 3,000 real images along with the \(5^{\text{th}}\) percentile.}
    \label{fig:sup_distances}
\end{figure}
\newpage

\subsection*{Validation of the similarity metrics}
\begin{figure}[h!]
    \centering
    \includegraphics[width=.9\linewidth]{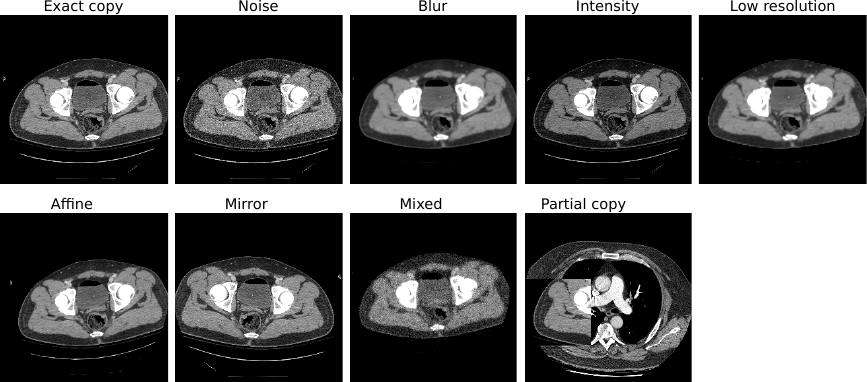}
    \caption{Examples of transforms used to test the similarity metrics.}
    \label{fig:augmentations}
\end{figure}

To evaluate the similarity metrics used in the main analysis, we built a reference database from 60 patients (16{,}279 images) of the RSNA 2020 PE dataset. \emph{Copies} were 1{,}000 images drawn from this database, each modified by one of nine transformations: exact copy (unchanged), noise (Gaussian noise), blur (Gaussian blur), intensity (intensity change), low resolution (downsampled then upsampled), affine (rotation $\leq 7^\circ$, scaling $\pm 7\%$, shift $\leq 10$\,px), mirror (left--right flip), mixed (all transforms combined), and partial copy (a patch from another image pasted in). These augmentations are depicted in Fig.~\ref{fig:augmentations}. \emph{Non-copies} were 1{,}000 images from 30 further patients, not part of the reference database, modified the same way. For each of these 2{,}000 query images, we searched the full reference database (60 patients, 16{,}279 images) for its most similar image.

We evaluated this search in two ways. First, for the \emph{copies}, we checked whether their source image was retrieved, measured as recall@1 (R@1), counting a match within $\pm2$ images of the source as correct since neighboring images are nearly identical. Second, we checked whether \emph{copies} scored higher than \emph{non-copies}, using each query's top-1 similarity to the reference database as its score; separability was summarized by the area under the ROC curve (AUROC), computed separately for each metric and copy type from 1{,}000 copies and 1{,}000 non-copies. \\
Table~\ref{tab:checkSimilarity} shows the results. Pixel correlation perfectly retrieves the source image (R@1\,=\,1.00) for copies that preserve the original pixel layout (exact copy through low resolution), but drops sharply once the image is geometrically altered (R@1\,=\,0.23--0.31 for affine, mirror, and mixed). SSCD remains robust across all these transformations (R@1\,$\geq$\,0.99), including the geometric ones. Both metrics fail to retrieve partial copies (R@1\,=\,0.01 and 0.12, respectively), where only a small region is shared with the source image.

\newpage

\begin{table*}[t]
\centering
\caption{Copy-detection benchmark for the two similarity metrics used in the main analysis. For each copy type, 1{,}000 copies and 1{,}000 non-copies (queries) were compared against a reference database of 16{,}279 real CT images. R@1: fraction of copies whose nearest database match is their own source image (within $\pm$2 images of the same series). AUROC: how well each metric's top-1 similarity score separates copies from non-copies (0.5 = chance). Copies / Non-copies: mean $\pm$ SD of the top-1 similarity score for copies and non-copies, respectively.}
\small
\setlength{\tabcolsep}{3pt}
\begin{tabular}{l cccc cccc}
\toprule
 & \multicolumn{4}{c}{Pixel correlation} & \multicolumn{4}{c}{SSCD} \\
\cmidrule(lr){2-5} \cmidrule(lr){6-9}
Copy type & R@1 & AUROC & Copies & Non-copies & R@1 & AUROC & Copies & Non-copies \\
\midrule
Exact copy & 1.00 & 1.00 & 1.00 $\pm$ 0.00 & 0.85 $\pm$ 0.06 & 1.00 & 1.00 & 1.00 $\pm$ 0.00 & 0.70 $\pm$ 0.05 \\
Noise & 1.00 & 1.00 & 1.00 $\pm$ 0.00 & 0.85 $\pm$ 0.06 & 1.00 & 1.00 & 0.93 $\pm$ 0.02 & 0.67 $\pm$ 0.05 \\
Blur & 1.00 & 1.00 & 1.00 $\pm$ 0.00 & 0.85 $\pm$ 0.06 & 1.00 & 1.00 & 0.92 $\pm$ 0.03 & 0.66 $\pm$ 0.05 \\
Intensity & 1.00 & 1.00 & 1.00 $\pm$ 0.00 & 0.85 $\pm$ 0.06 & 1.00 & 1.00 & 0.99 $\pm$ 0.01 & 0.69 $\pm$ 0.05 \\
Low resolution & 1.00 & 1.00 & 1.00 $\pm$ 0.00 & 0.85 $\pm$ 0.06 & 1.00 & 1.00 & 0.90 $\pm$ 0.03 & 0.65 $\pm$ 0.05 \\
Affine & 0.31 & 0.67 & 0.85 $\pm$ 0.06 & 0.81 $\pm$ 0.07 & 1.00 & 1.00 & 0.92 $\pm$ 0.02 & 0.68 $\pm$ 0.05 \\
Mirror & 0.23 & 0.61 & 0.86 $\pm$ 0.06 & 0.83 $\pm$ 0.07 & 1.00 & 1.00 & 0.96 $\pm$ 0.01 & 0.69 $\pm$ 0.05 \\
Mixed & 0.30 & 0.69 & 0.86 $\pm$ 0.05 & 0.81 $\pm$ 0.07 & 0.99 & 1.00 & 0.79 $\pm$ 0.03 & 0.61 $\pm$ 0.04 \\
Partial copy & 0.01 & 0.50 & 0.82 $\pm$ 0.06 & 0.82 $\pm$ 0.06 & 0.12 & 0.49 & 0.63 $\pm$ 0.06 & 0.63 $\pm$ 0.06 \\
\bottomrule
\end{tabular}
\label{tab:checkSimilarity}
\end{table*}
\clearpage
\subsection*{Ablation Study}
\begin{figure}[h!]
    \centering
    \includegraphics[width=1.\linewidth]{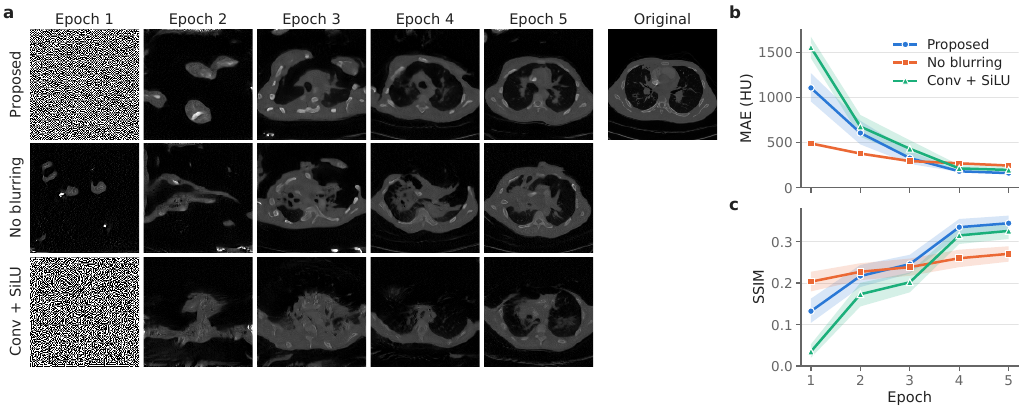}
\caption{Ablation of blurring diffusion and the gated residual block during the first five training epochs (128$\times$128 models). \textbf{a} A held-out PE-positive slice (original, top right) encoded into its semantic code and decoded at each epoch by the proposed model, a model without blurring, and a model with conv + SiLU instead of the gate. All panels start from identical noise (window $-1000$ to $3000$\,HU). \textbf{b} Mean absolute error (MAE) and \textbf{c} structural similarity (SSIM, mediastinal window) between decoded and original slices, averaged over 100 held-out slices; shaded bands show 95\% bootstrap confidence intervals.}
    \label{fig:trainingAblation}
\end{figure}

We trained two ablated variants of the 128$\times$128 model alongside the proposed model, each differing in a single component: one used standard instead of blurring diffusion, the other replaced the gated residual block with a convolution followed by SiLU. Data, initial weights, optimizer, batch order and random seeds were identical. After every epoch, 100 held-out slices were encoded and decoded from identical noise, and each decoded slice was compared with its original using MAE and SSIM (Fig. \ref{fig:trainingAblation}).
Without blurring diffusion, the model produced non-noise images already after the first epoch, ahead of the proposed model at this stage, but took longer to recover the correct anatomical outline: for example, the body contour was still not fully closed after epoch 4. This is also reflected in the metrics: MAE was lower at epochs 1--3 but then plateaued, and by epoch 5 this variant showed the largest reconstruction error (MAE 244 vs.\ 160\,HU; SSIM 0.27 vs.\ 0.34; both $p<0.001$, Wilcoxon signed-rank test). The conv~+~SiLU variant followed a similar trajectory to the proposed model but still showed a higher reconstruction error at epoch 5, producing less anatomically plausible images (MAE 195\,HU, SSIM 0.33; both $p<0.001$). Both components therefore contribute to a more faithful reconstruction of a slice from its semantic code, with blurring diffusion having the larger effect.

\newpage

\subsection*{Hyperparameters of the conditional baseline}

\begin{figure}[h!]
    \centering
    \includegraphics[width=1.\linewidth]{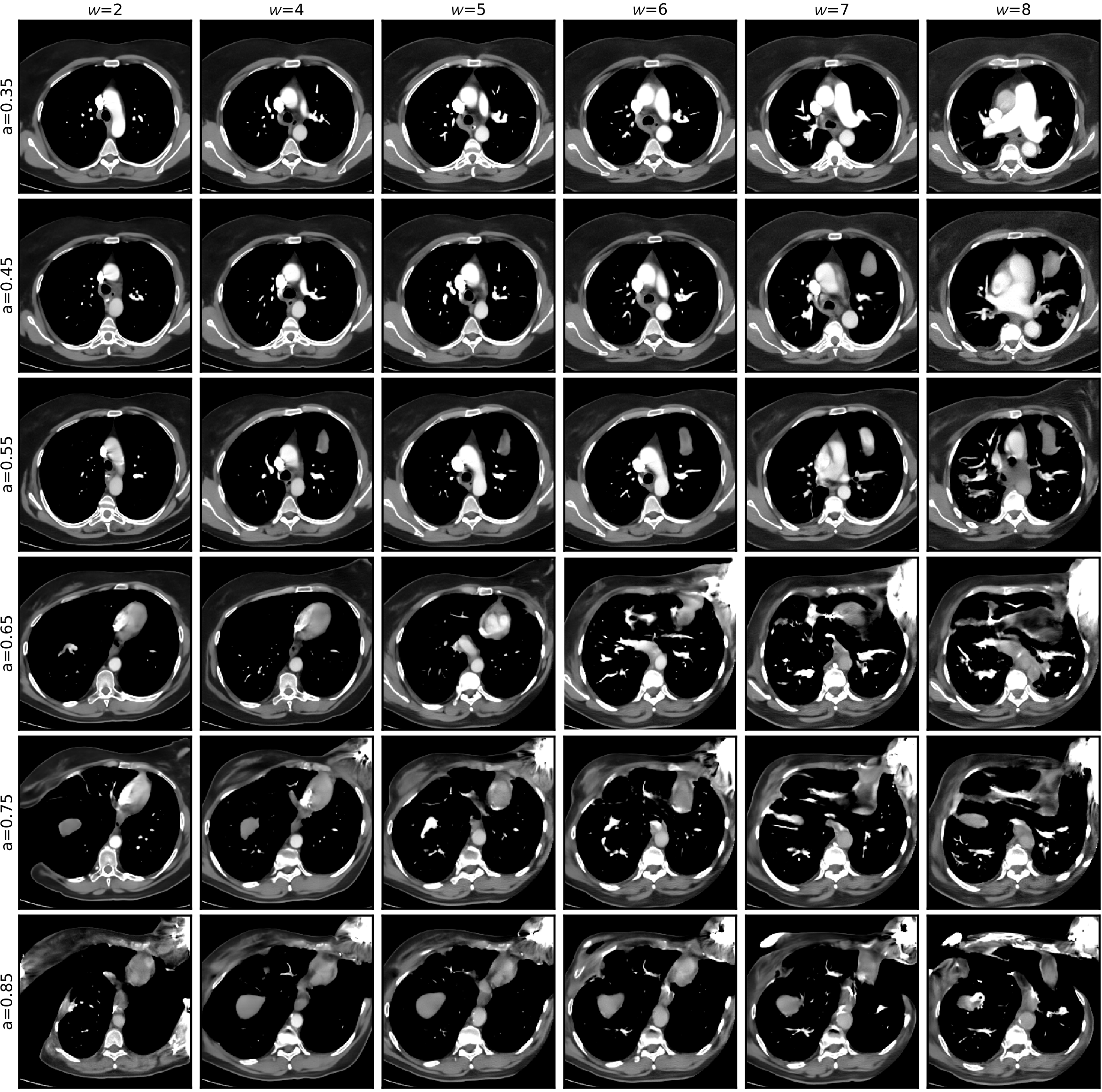}
\caption{Effect of the classifier-free guidance scale $w$ and noise retention parameter $a$ on the image generation process of the conditional diffusion baseline model.}
    \label{fig:condBaselineHyperparameters}
\end{figure}

\end{document}